\documentclass[journal=jpccck,manuscript=article]{achemso}

\usepackage[version=3]{mhchem} 

\let\oldAA\AA
\renewcommand{\AA}{\text{\normalfont\oldAA}}

\author{Freerk Sch\"utt}
\affiliation{Carl-von-Ossietzky Universit{\"a}t Oldenburg, Institute of Physics, 26129 Oldenburg, Germany}
\author{Ana M. Valencia}
\affiliation{Carl-von-Ossietzky Universit{\"a}t Oldenburg, Institute of Physics, 26129 Oldenburg, Germany}
\altaffiliation{Humboldt-Universit\"at zu Berlin, Physics Department and IRIS Adlershof, 12489 Berlin, Germany}
\author{Caterina Cocchi}
\affiliation{Carl-von-Ossietzky Universit{\"a}t Oldenburg, Institute of Physics, 26129 Oldenburg, Germany}
\alsoaffiliation{Carl-von-Ossietzky Universit{\"a}t Oldenburg, Center for Nanoscale Dynamics, 26129 Oldenburg, Germany}
\altaffiliation{Humboldt-Universit\"at zu Berlin, Physics Department and IRIS Adlershof, 12489 Berlin, Germany}
\email{caterina.cocchi@uni-oldenburg.de}

\title{Electronic Structure and Optical Properties of Tin Iodide Solution Complexes}

\begin{document}

\begin{tocentry}
\begin{figure}[H]
    \centering
    \includegraphics[width=8.25 cm]{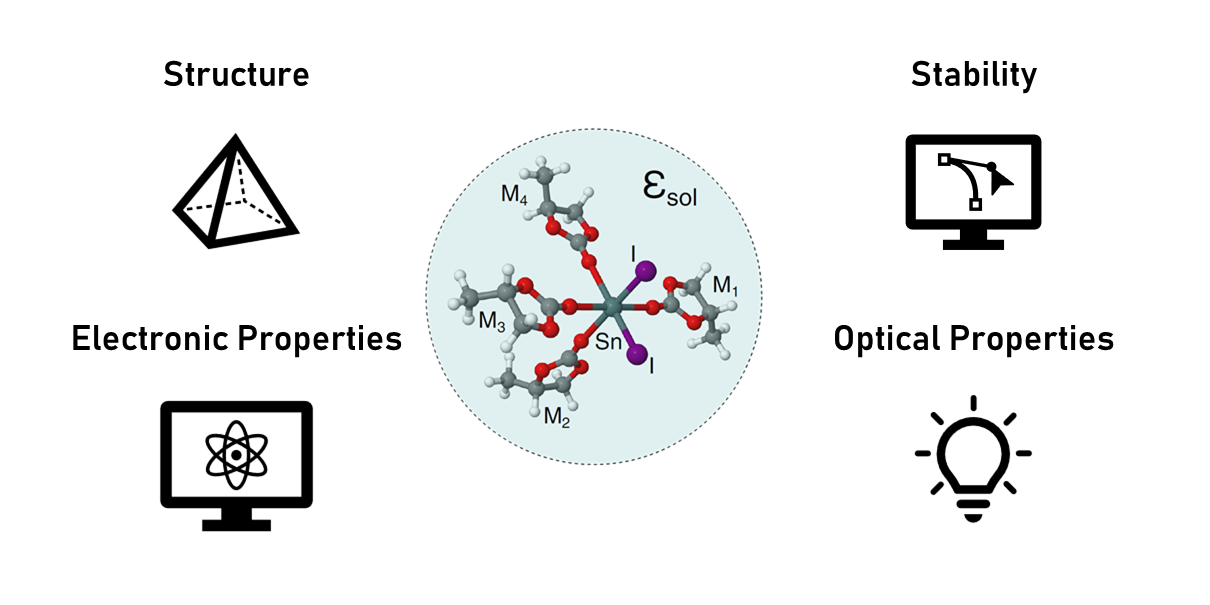}
\end{figure}

\end{tocentry}

\newpage
\begin{abstract}
The emerging interest in tin-halide perovskites demands a robust understanding of the fundamental properties of these materials starting from the earliest steps of their synthesis. In a first-principles work based on time-dependent density-functional theory, we investigate the structural, energetic, electronic, and optical properties of 14 tin-iodide solution complexes formed by the \ce{SnI2} unit tetracoordinated with molecules of common solvents, which we classify according to their Gutmann's donor number. We find that all considered complexes are energetically stable and their formation energy expectedly increases with the donating ability of the solvent. The energies of the frontier states are affected by the choice of the solvent with their absolute values decreasing with the donor number. The occupied orbitals are predominantly localized on the tin-iodide unit while the unoccupied ones are distributed also on the solvent molecules. Owing to this partial wave-function overlap, the first optical excitation is generally weak, although the spectral weight is red-shifted by the solvent molecules being coordinated to \ce{SnI2} in comparison to the reference obtained for this molecule alone. Comparisons with results obtained on the same level of theory on Pb-based counterparts corroborate our analysis. The outcomes of this study provide quantum-mechanical insight into the fundamental properties of tin-iodide solution complexes. This knowledge is valuable in the research on lead-free halide perovskites and their precursors.
\end{abstract}

\newpage
\section{Introduction}
Tin halide perovskites are emerging materials for photovoltaic and optoelectronic applications~\cite{jiang2021tin,konstantakou-stergiopoulos2017critical,nasti-abate2020tin} responding to the pressing quest for non-polluting and non-toxic compounds for solar cells~\cite{aina-etal2021earthabundant,ikram-etal2022recent}.
Like their Pb-based counterparts, these systems are produced via solution chemistry and the efficiency of the resulting thin films crucially depends on the physico-chemical properties of the precursors~\cite{digirolamo-etal2021solvents,jiang-etal2021onestep,pascual-etal2022lights}.
The choice of the solvent is a particularly critical point, as it can induce oxidation~\cite{leijtens-etal2017mechanism,pascual-etal2020origin,radicchi2022solvent} and thus contribute to the rapid degradation of the material in operational conditions~\cite{digirolamo-etal2021solvents,li-etal2021ionic,meng-etal2020highly}.
As a result of these efforts, it is now common knowledge that the popular solvent dimethyl sulfoxide (DMSO) is detrimental for the stability of tin halide perovskite solar cells~\cite{digirolamo-etal2021solvents,pascual-etal2022lights}.
This finding has stimulated the search for guiding principles toward the choice of optimal solvents for the synthesis of this class of materials~\cite{gu-etal2020robotbased,manion-etal2020highthroughput}.

The missing piece of the puzzle is a systematic, first-principles analysis of the microscopic properties of tin halide solution complexes as building blocks for corresponding perovskite structures.
Similar studies performed on lead halide counterparts have significantly contributed to disclosing the characteristics of these compounds~\cite{kaiser-etal2021iodide,procida-etal2021firstprinciples,radicchi-etal2019understanding,radicchi2020structural,schier-etal2021formation,valencia-etal2021optical} and, hence, to better understanding the evolution process of these systems from solution precursors to thin films through several intermediate steps~\cite{hamill-etal2018influence,huang-etal2022intermediate}. 
In the case of Sn-based halide perovskites, such efforts have been conducted primarily in conjunction with experiments~\cite{digirolamo-etal2021solvents}. 
A dedicated \textit{ab initio} study is, to the best of our knowledge, still missing.

In the framework of time-dependent density functional theory coupled to the polarizable continuum model, we investigate the physical properties of 14 solution complexes, \ce{SnI2M4}, where $M$ indicates a common solvent molecule chosen among compounds previously adopted experimentally in the synthesis of tin-halide precursors~\cite {digirolamo-etal2021solvents}.
In the presented analysis, we focus on the structural, energetic, electronic, and optical properties of these systems.
We characterize the considered compounds by looking at bond lengths and angles and by comparing them with their counterparts in \ce{PbI2M4} complexes. We assess their formation energy and rationalize our findings with respect to the donor number of the solvent. Finally, we inspect the electronic structure in order to interpret their absorption spectra and disclose the characteristics of their optical excitations, including their composition and spatial distribution among the constituents.

\section{Computational Methods}
All \textit{ab initio} calculations presented in this study are performed using density functional theory (DFT)~\cite{hohenberg-kohn1964inhomogeneous, kohn-sham1965selfconsistent} as implemented in the Gaussian~16 software package~\cite{g16}.
The implicit solvent interactions are simulated with the polarizable continuum model (PCM)~ \cite{miertus-etal1981electrostatic, amovilli-etal1998recent} adopting the values for the dielectric constants of the solvents listed in Figure~\ref{fig:solvent-structure-overview}. 
Van der Waals interactions are accounted for using the semi-empirical Grimme D3 dispersion scheme~\cite{grimme-etal2010consistent}.
Reference calculations on the isolated \ce{SnI2} complex assumed to be implicitly solvated in DMSO are performed adopting the PCM.

The geometries of the $\text{SnI}_2\text{M}_4$ complexes are optimized through the minimization of the interatomic forces. By means of additional frequency calculations, we checked that all the obtained structures represent global minima except for \ce{SnI2(TMU)4} and \ce{SnI2(HMPA)4}, which exhibit one and four negative frequencies, respectively. We argue that this result does not impact the following analysis of the electronic and optical properties of the complexes, which is primarily focused on identifying trends.
For these optimizations, we used the generalized gradient approximation in the Perdew-Burke-Ernzerhof (PBE) parameterization~\cite{perdew-etal1996generalized} together with the LANL2DZ basis set; pseudopotentials are adopted for Sn and I atoms, and the double-zeta basis set cc-pVDZ for the remaining lighter species.
We checked that the choice of this functional, which is substantially more efficient than hybrid approximations and adopted in previous studies on similar systems~\cite{kaiser-etal2021iodide,procida-etal2021firstprinciples,radicchi-etal2019understanding,schier-etal2021formation,valencia-etal2021optical}, does not affect the electronic structure of the resulting optimized structures.
For single-point-geometry DFT calculations, for the population and natural bond order (NBO) analysis, as well as for the time-dependent DFT (TDDFT) runs, we adopt the range-separated hybrid functional CAM-B3LYP~\cite{yanai-etal2004new} in combination with the SDD basis set together with the corresponding SDD pseudopotentials for Sn and I, and the quadruple-zeta cc-pVQZ basis set for the lighter atoms.
The output of (TD)DFT calculations is post-processed to obtain relevant information on the systems.
The molecular orbitals (MOs) of the \ce{SnI2M4} complexes are analyzed with the natural atomic orbital (NAO) method as implemented in the \texttt{Multiwfn}~\cite{lu-chen2012multiwfn} software on the basis of the NBO calculations performed with Gaussian~16~\cite{lu-chen2013bond,zhong-etal2018molecular}.

The properties of the complexes are analyzed in light of the Gutmann donor number ($D_N$) of the solvents as the primary classifier according to their Lewis basicity. This property is recognized as a determining factor for the effectiveness of the solvents and specifically for their ability to solvate halide perovskite precursors~\cite{abdel-shakour-etal2021highefficiency,hamill-etal2018influence,stevenson-etal2017mayer,romiluyi-etal2021efficacy}.
The values of $D_N$ adopted in this work are taken from available references and, in the absence of the latter, estimated from first principles from the ionization potential (IP) and the electron affinity (EA) of the compounds as computed from DFT~\cite{miranda-quintana-smiatek2021calculation} (see Figure~\ref{fig:solvent-structure-overview} and Table~S1 in the Supporting Information).
The absorption spectra evaluated from TDDFT+PCM are plotted including a Lorentzian broadening of 70~meV for the excitation linewidths.

\section{Systems}
\begin{figure}
    \centering
    \includegraphics[width=\textwidth]{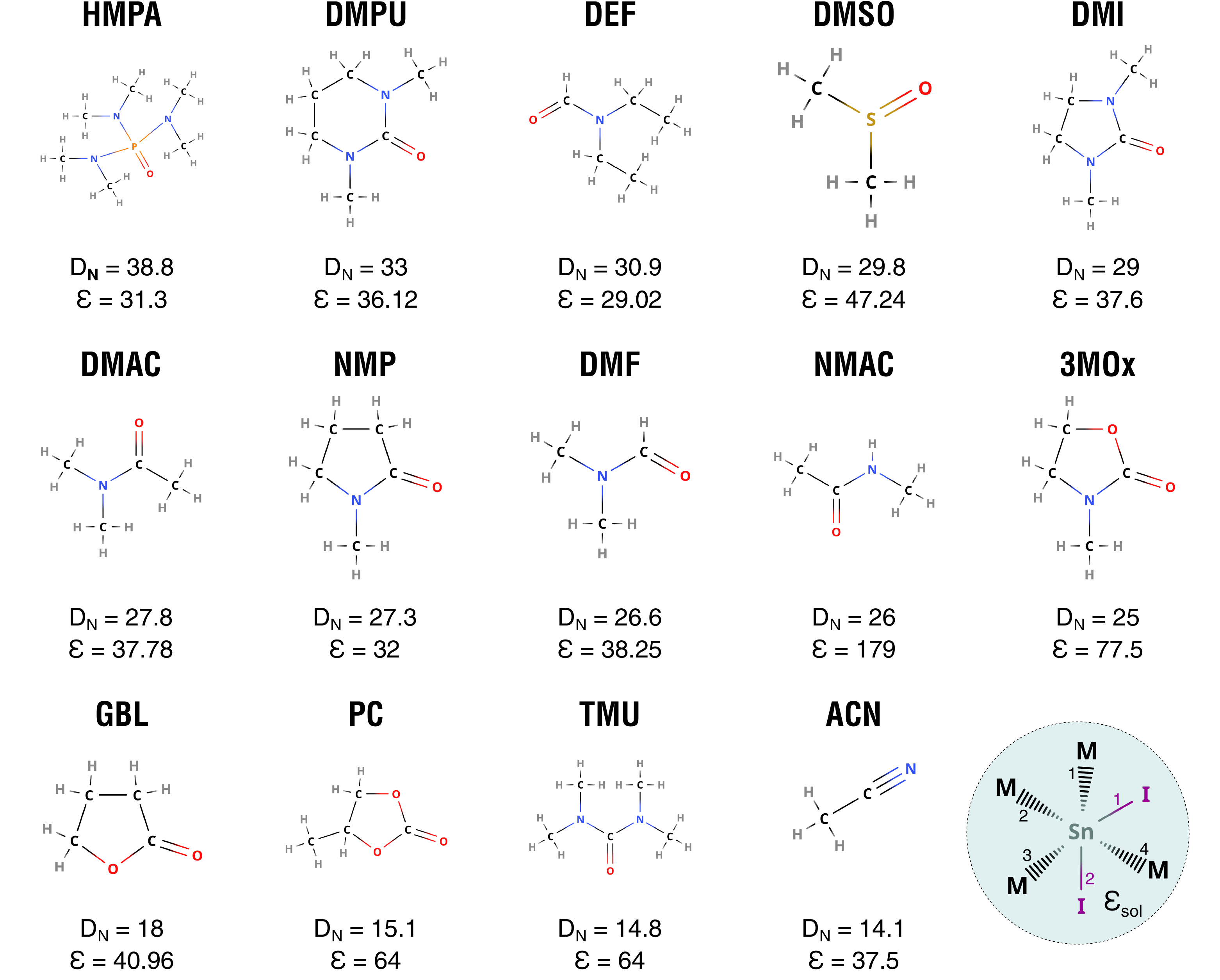}
    \caption{Overview of the 14 solvents considered in this study with their respective Gutmann's donor number ($D_N$) and static dielectric constant ($\varepsilon$): Hexamethylphosphoramide (HMPA), N, N'-Dimethylpropyleneurea (DMPU), N,N-Diethylformamide (DEF), Dimethyl sulfoxide (DMSO), 1,3-Dimethyl-2-imidazolidinone (DMI), Dimethylacetamide (DMAC), N-Methyl-2-pyrrolidone (NMP), Dimethylformamide (DMF), N-Methylacetamide (NMAC), 3-Methyl-2-oxazolidinone (3MOx), $\gamma$-Butyrolactone (GBL), propylene carbonate (PC), Tetramethylurea (TMU), and acetonitrile (ACN). On the bottom right panel, a sketch of the \ce{SnI2M4} compounds in the solvent cavity is reported: the numbers label the Sn-I (purple) and the Sn-M (black) bonds.}
    \label{fig:solvent-structure-overview}
\end{figure}

In this study, we consider 14 tin iodide solution complexes characterized by a \ce{SnI2} backbone in the axial configuration surrounded by four solvent molecules binding to it (see Figure~\ref{fig:solvent-structure-overview}). 
The initial geometries are set up based on earlier works on lead iodide solution complexes~\cite{schier-etal2021formation,procida-etal2021firstprinciples,digirolamo-etal2021solvents} while the choice of solvents is based on experimental work on tin halide perovskite precursors~\cite{digirolamo-etal2021solvents}.
We consider the common solvents propylene carbonate (PC), which is a green compound, and DMSO as a point of reference for comparisons with Pb-based counterparts. In fact, DMSO leads to the oxidation of Sn$^{2+}$ and, as such, is detrimental to tin-halide film formation~\cite{pascual-etal2020origin,saidaminov-etal2020conventional}.
Other chosen solvents belong to the class of amide (DEF, DMF, DMAC, NMAC, NMP) and diamide (TMU, DMI, and DMPU).
The remaining ones are representative of carbamate (3MOx), of solvents containing the P=O bond (HMPA), and of species dissolving from formaldiminium iodide (GBL).
Finally, we consider acetonitrile (ACN) which binds to \ce{SnI2} via a nitrogen atoms in contrast to all the other solvents mentioned above which bind to metal-halide unit via an oxygen (see Figure~\ref{fig:solvent-structure-overview}).
The \ce{SnI2M4} compounds are simulated atomistically in order to assess the quantum-mechanical interactions holding them together.
The additional electrostatic forces coming from the surrounding solvent medium are accounted for implicitly within the PCM cavity (see Figure~\ref{fig:solvent-structure-overview}, bottom right panel).

\section{Results and Discussion}

\subsection{Structural Properties}
\begin{figure}[h!]
    \centering
    \includegraphics[width=0.5\textwidth]{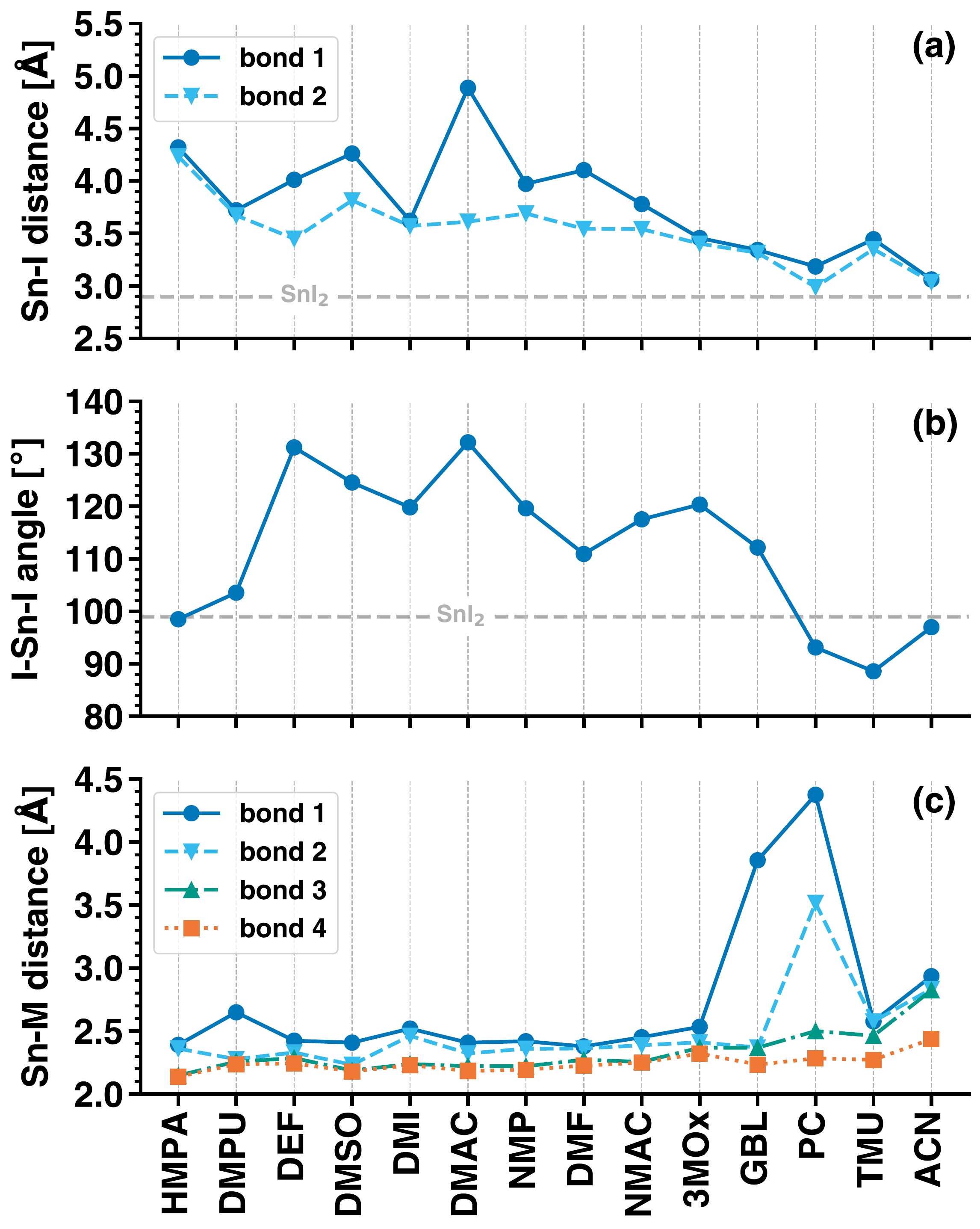}
    \caption{Structural properties of the \ce{SnI2M4} complexes, including (a) the Sn-I distances, (b) the I-Sn-I angle, and (c) the distances between Sn and the anchoring atoms solvent molecule M (Sn-M distance). The solvents are ordered on the $x$-axis with respect to their donor number $D_N$, decreasing from left to right. The horizontal dashed line in panels (a) and (b) mark the reference value obtained for \ce{SnI2}.}
    \label{fig:structuralproperties}
\end{figure}

We start our analysis from the structural properties of the 14 \ce{SnI2M4} complexes considered in this work.
Specifically, we inspect the distances between Sn and I atoms in the metal-halide backbone, the I-Sn-I angle, and the separation between the Sn atoms and the solvent molecules.
Regardless of the solvent, we notice at a glance that the Sn-I bond length is systematically increased in the solution complexes with respect to the \ce{SnI2} molecule, see Figure~\ref{fig:structuralproperties}a.
The bond elongation is enhanced by solvents with a large donor number; in fact, in the presence of weakly donating solvents such as PC or ACN, the Sn-I distances remain very close to the reference values obtained for \ce{SnI2}.
In some cases, the interactions with solvent molecules induce asymmetric distortions in the tin-iodide backbone. 
In particular, the result obtained with DMAC sticks out.
In this case, the \ce{SnI2} unit is significantly stretched, with one I atom at approximately 3.6~\AA{} apart from Sn and the other one at almost 5~\AA{}. 
In this configuration, the Sn and I atoms interact only weakly with arguably only one chemical bond remaining in place. As we will see in the following, this characteristic has remarkable consequences on the electronic and optical properties of the complex.

Inspection of the trends for the I-Sn-I angle provides additional indications on the structural changes induced by the solvent molecules on \ce{SnI2} (see Figure~\ref{fig:structuralproperties}b). 
Despite the elongation of the Sn-I bond length, interaction with HMPA does not lead to any variation of the I-Sn-I angle.
The presence of DMPU, ACN, and PC does not strongly modify this structural parameter either. 
On the other hand, the increase in the Sn-I separation is accompanied by a sizeable increase in the I-Sn-I angle in presence of solvents with intermediate values of $D_N$.
The angle tends to close up with weakly donating solvents such as PC, TMU, and ACN, although variations above 10$^{\circ}$ with respect to the reference value in \ce{SnI2} can be appreciated only with TMU.

Finally, we analyze the separations between tin atoms in \ce{SnI2} and the surrounding solvent molecules.
As shown in Figure~\ref{fig:structuralproperties}c, in most cases, the four Sn-M distances within a complex are almost equal, ranging from 2.1~\AA{} to 2.6~\AA{}. 
Notice that the average of these values, namely 2.3~\AA{}, is very close to the tabulated experimental value for the Sn-O bond length (2.2~\AA{})~\cite{moore-pauling1941crystal} as well as to DFT results obtained for the same quantity in tin oxide~\cite{pedersen-luisier2014lithiation}.
Yet, a couple of outliers can be identified: Sn-O distances above 3.5~\AA{} testify that \ce{SnI2} does not bind with one GBL molecule and with two PC ones. 
In \ce{SnI2(PC)4}, we ascribe it to the weak ability of the solvent to bind to tin iodide, where, indeed, the Sn-I separation remains almost intact with respect to the isolated \ce{SnI2} unit (Figure~\ref{fig:structuralproperties}a). 
Finally, the larger Sn-M distances found in \ce{SnI2(ACN)4} are likely the result of the reduced binding ability of ACN in comparison with solvents with higher $D_N$ that are anchored to \ce{SnI2} via an oxygen atom. 
In fact, reference DFT values for Sn-N distances are between 2.2 and 2.3~\AA{}~\cite{kim-etal2018novel}, namely very similar to Sn-O separations.

Overall, the structural trends displayed in Figure~\ref{fig:structuralproperties} can be rationalized with the coordinating ability of the solvents.
Compounds with high $D_N$ values strongly coordinate with the metal-halide salts, competing against \ce{I^-} for the coordination sites around the central metal cation. 
This is reflected in the larger distances between the metallic cation (Sn$^{2+}$) and the \ce{I^-} anions and smaller distances between Sn$^{2+}$ and the solvent molecules.
On the other hand, solvents with lower $D_N$ have a weaker coordination ability with the cation, leading to smaller Sn-I distances and larger Sn-M separations~\cite{hamill-etal2018influence,romiluyi-etal2021efficacy}.

\begin{figure}[h!]
    \centering
    \includegraphics[width=0.5\textwidth]{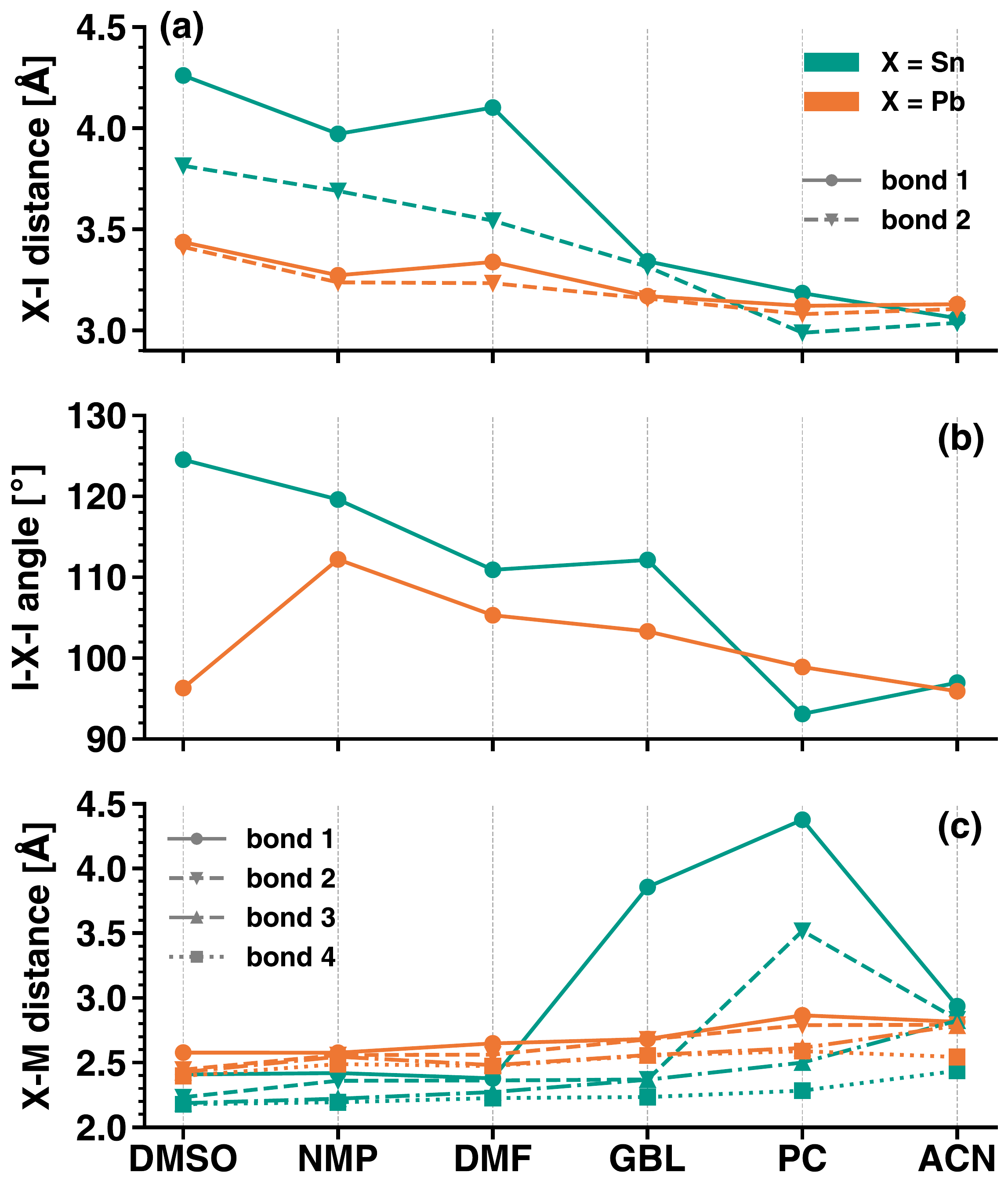}
    \caption{Structural properties of six \ce{SnI2M4} complexes compared to their \ce{PbI2M4} counterparts calculated with identical computational parameters in Ref.~\citenum{schier-etal2021formation}: (a) Metal-iodine (X-I) distances, (b) metal-iodide (I-X-I) angle, and (c) metal-solvent (X-M) distances are shown. Solvents with decreasing donor number are displayed on the $x$-axis from left to right. }
    \label{fig:structuralpropertiescomparison}
\end{figure}

The structural properties of six \ce{SnI2M4} complexes considered in this work can be directly compared to their \ce{PbI2M4} counterparts previously investigated on the same level of theory~\cite{schier-etal2021formation}.
The above-discussed trends of increasing metal-I distance and decreasing I-metal-I angle and metal-M distances for decreasing $D_N$ are common to both sets of complexes (see Figure~\ref{fig:structuralpropertiescomparison}).
In the presence of solvents with high $D_N$, such as DMSO, NMP, and to some extent DMF, Sn-I distances are larger than their Pb-I counterparts by more than 0.5~\AA{}.
Notice that the distance from I is almost identical for Sn and Pb in the \ce{SnI2} and \ce{PbI2} molecules, respectively, see Figure~\ref{fig:structuralproperties}a and Ref.~\citenum{schier-etal2021formation}.
With decreasing solvent donor numbers, Sn-I and Pb-I distances approach each other until they become almost identical with the solvents PC and ACN.
Overall we notice more symmetric structures in the Pb-based complexes, as testified by the very similar Pb-I bond lengths regardless of the solvents (see Figure~\ref{fig:structuralpropertiescomparison}a).
In contrast, Sn coordinates differently with the two I atoms likely due to its smaller size compared to both Pb and iodine itself. 
This characteristic can be associated with the tendency to lattice distortion exhibited by bulk tin-based halide perovskites~\cite{chung-etal2012cssni3}.

Moving on to the angles of the metal-iodide backbone, we notice that those in the tin-based complexes are systematically larger than those in the Pb-based compounds; the only exception is for the compounds formed with PC (Figure~\ref{fig:structuralpropertiescomparison}b).
We correlate this trend with the one discussed above for the metal-iodine distances: large bond lengths lead to an opening of the angle and vice versa.
Lastly, the Pb-M distances are consistently larger than the Sn-M distances by approximately 0.2~\AA{} and are also more homogeneous among each other (Figure~\ref{fig:structuralpropertiescomparison}c). 
In contrast to \ce{SnI2(GBL)4} and \ce{SnI2(PC)4}, where one and two molecules, respectively, do not bind to the metal-iodide center, in the Pb-based counterparts all solvents are coordinated with \ce{PbI2}~\cite{schier-etal2021formation}.
In general, the choice of the solvent seems to have a larger influence on the structure of the Sn-based complexes compared to the Pb-based ones. 
The stronger solvent-metal coordination and subsequent weakening of the metal-I interaction appear more significant in Sn-based complexes compared to the Pb-based ones.
The structural variability of tin-iodide complexes can be ascribed to the small size of Sn compared to Pb, which likely has a more stabilizing influence as the central cation and shows generally stronger coordination. 

We conclude this analysis by commenting that the \ce{SnI2M4} complex that is structurally most similar to its \ce{PbI2M4} counterpart is the one including ACN (Figure~\ref{fig:structuralpropertiescomparison}).
We interpret this behavior based on the weakest ability of this solvent to coordinate with the metal iodide center in comparison with the other ones considered in this work. 
In contrast, the largest structural differences are seen between \ce{SnI2(DMSO)4} and \ce{PbI2(DMSO)4}.
This result is in line with recent findings relating structural distortions occurring in Sn-halide perovskites synthesized with DMSO, due to solvent-induced oxidation of tin~\cite{digirolamo-etal2021solvents, pascual-etal2020origin, pascual-etal2022lights,saidaminov-etal2020conventional}.


\subsection{Energetic Stability}
With the insight gained from the analysis of the structural properties, we move on to the investigation of the energetic stability of the \ce{SnI2M4} complexes.
The formation energy, $E_f$, is calculated as
\begin{equation}
    \label{Eq:formationenergy}
    E_{f} = \frac14(E_\text{tot} - E_{\ce{SnI2}} - E_{\text{M}_{1}} - E_{\text{M}_{2}} - E_{\text{M}_{3}} - E_{\text{M}_{4}}),
\end{equation}
where $E_\text{tot}$ is the total energy of the complex in its optimized structure, $E_{\ce{SnI2}}$ is the energy of the $\ce{SnI2}$ molecule in the solution complex, and $E_{\text{M}_{i}}$ the energy of each of the four solvent molecules in their relaxed geometry within the complex~\cite{schier-etal2021formation}.
In Eq.~\eqref{Eq:formationenergy}, the energy contributions from all four solvent molecules are considered separately, taking into account the differences induced in their geometries when they bind to \ce{SnI2}.
Higher (lower) stabilities are indicated by more (less) negative values of $E_f$.

\begin{figure}[h]
    \centering
    \includegraphics[width=0.5\textwidth]{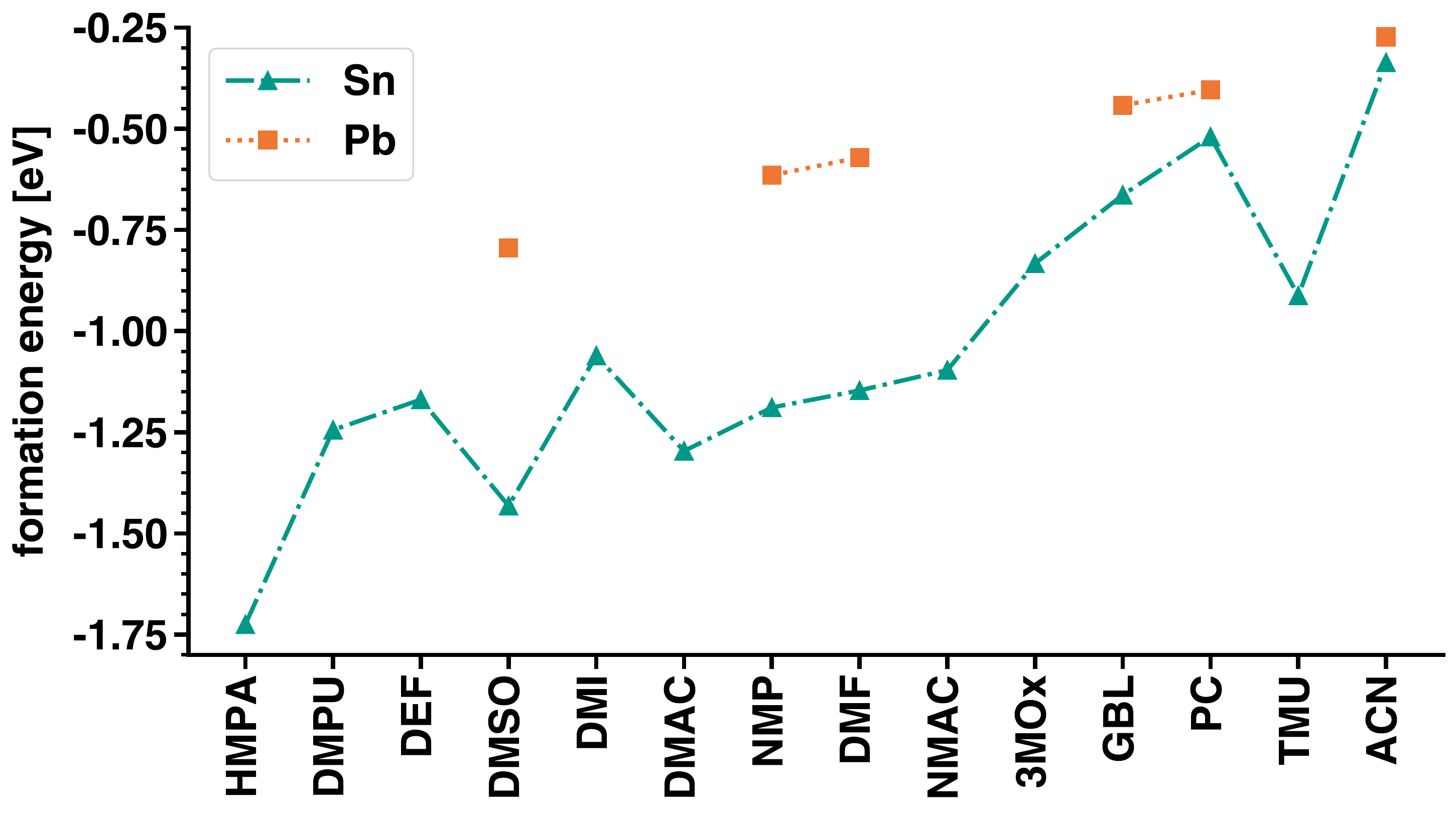}
    \caption{Formation energies of the 14 \ce{SnI2M4} complexes investigated in this work (green triangles); corresponding values for Pb-based counterparts taken from Ref.~\citenum{schier-etal2021formation} (orange squares) are reported for comparison. 
    }
    \label{fig:formationenergy}
\end{figure}

The results plotted in Figure~\ref{fig:formationenergy} (dataset available in Table~S3) indicate that all the considered complexes are stable.
The largest formation energies are found in the presence of the HMPA solvent which contains the P=O group.
The least stable complex, on the other hand, is the one formed with ACN, the only solvent molecule included in our analysis that is bound to tin-iodide through an N atom.
In general, we notice a clear tendency toward decreasing stability with decreasing $D_N$, in line with previous studies~\cite{digirolamo-etal2021solvents,hamill-etal2018influence,heo-etal2021enhancing}.
Quantitative and even qualitative differences between our results and those reported in Ref.~\citenum{digirolamo-etal2021solvents} for the same tin-iodide complexes can be attributed to different definitions adopted for the formation energy as well as to the chosen computational setups.  

We compare the formation energies displayed in Figure~\ref{fig:formationenergy} with the six counterparts computed for Pb-based analogs obtained with identical computational parameters~\cite{schier-etal2021formation}.
The decreasing stability with decreasing $D_N$ of the solvent discussed above is visible also in the lead-based complexes, where the trend is monotonic.
The values of $E_f$ computed for the Sn-based complexes are systematically more negative than those of the Pb-based ones~\cite{schier-etal2021formation}, possibly on account of the smaller size of Sn compared to Pb and thus its higher tendency to form (stable) chemical bonds.
In fact, in the presence of GBL and PC, whereby not all four molecules bind to \ce{SnI2}, formation energies are very close to the corresponding values for the Pb-based counterparts. 
The same holds true also for \ce{SnI2(ACN)4}, which exhibits a value of $E_f$ almost identical to the one obtained in Ref.~\citenum{schier-etal2021formation} for \ce{PbI2(ACN)4}.
Notice that the structural parameters of these two compounds are overlapping, too (see Figure~\ref{fig:structuralpropertiescomparison}).
In contrast, the results obtained for the tin-based complexes with DMSO, DMF, and NMP differ by several hundreds of meV from their counterparts with Pb (see Figure~\ref{fig:formationenergy}).
These trends can be rationalized again with the aid of the structural analysis reported in Figure~\ref{fig:structuralpropertiescomparison}.
In the \ce{XI2M4} complexes (with X=Sn, Pb and M=DMSO, NMP, DMF), the X-M distances are similar but the X-I separations differ substantially. 
This finding suggests an enhanced ionicity of the Sn-I bond compared to the Pb-I one in the presence of these high-$D_N$ solvents and a consequent stabilization of the \ce{SnI2M4} complexes through the binding with solvent molecules.

\subsection{Electronic Properties}

\begin{figure}[h]
    \centering
    \includegraphics[width=0.5\textwidth]{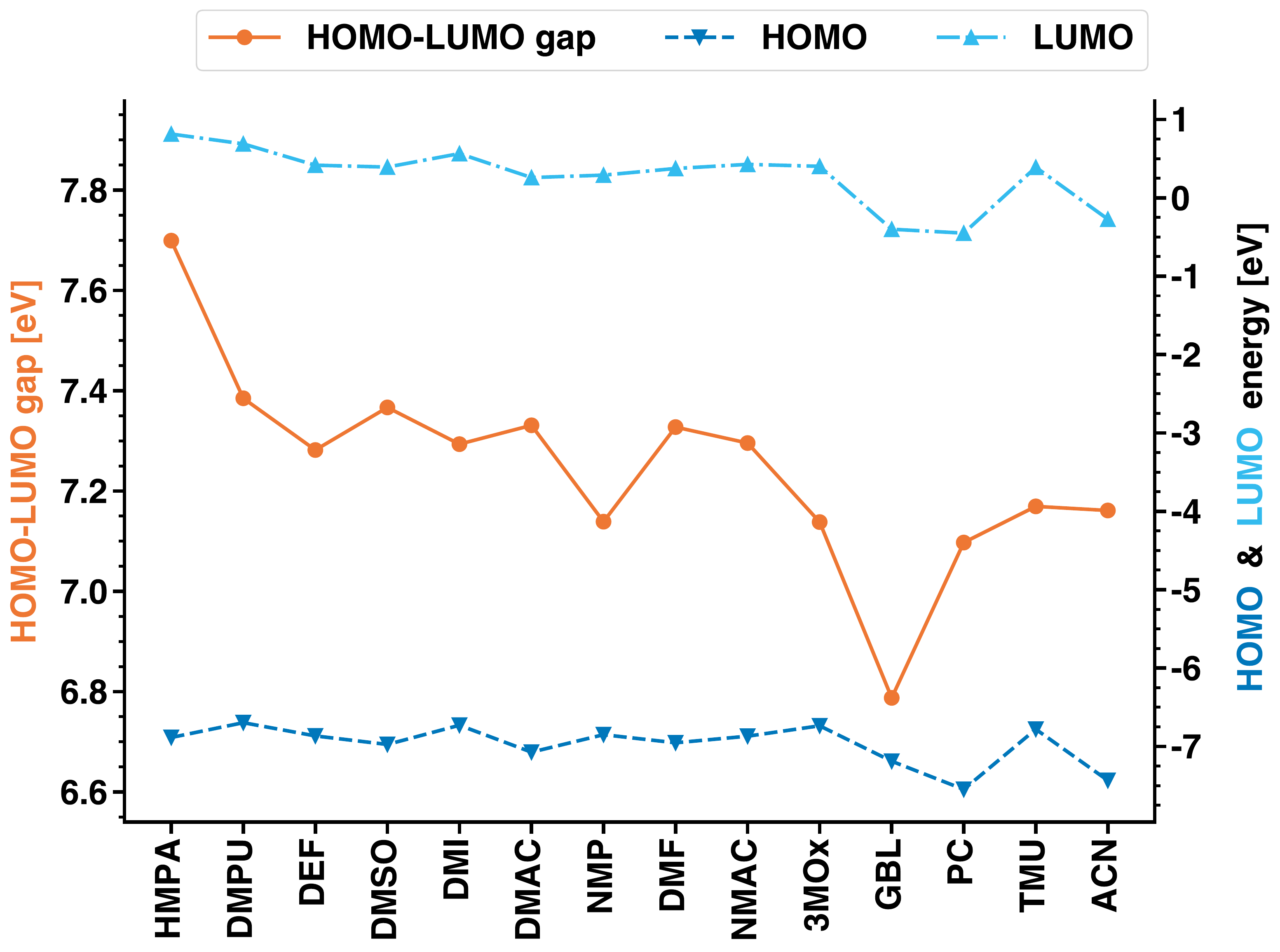}
    \caption{HOMO and LUMO energies of the 14 considered complexes plotted in dark and light blue, respectively, against the right $y$-axis, and the resulting HOMO-LUMO gap plotted in orange against the left $y$-axis.}
    \label{fig:h-l-gap}
\end{figure}

In the next step of our analysis, we focus on the electronic structure of the \ce{SnI2M4} complexes. 
We investigate the energy and the spatial distribution of the MOs, paying special attention to the highest-occupied and the lowest-unoccupied oribtals (HOMO and LUMO, respectively), and to the gap between them.
The results visualized in Figure~\ref{fig:h-l-gap} (data in Table~S4) suggest decreasing HOMO and LUMO energies for decreasing values of $D_N$, although this trend is not monotonic.
The electron-withdrawing or -donating ability of functional group covalently bound to the edges of carbon-based molecules was shown to tune the frontier levels upward or downward with respect to a reference value~\cite{cocchi-etal2011optical,cocchi-etal2012electronics}.
The involved chemistry here is different but the above-mentioned scenario can contribute to the understanding of the trends reported in Figure~\ref{fig:h-l-gap}.
Overall, the values of HOMO and LUMO energies reported for all the considered solvents vary within a window of about 1~eV, which is reflected also in their differences. 
The largest HOMO-LUMO gap, equal to 7.7 eV, pertains to \ce{SnI2(HMPA)4} while the smallest one (6.8 eV) to \ce{SnI2(GBL)4}.
The gap variations in these tin-based complexes are mainly ascribed to the fluctuating HOMO energies in contrast with the behavior of the Pb-based counterparts, where the shifts in the LUMO energies play a more crucial role~\cite{procida-etal2021firstprinciples}.

\begin{figure}[h]
    \centering
    \includegraphics[width=0.5\textwidth]{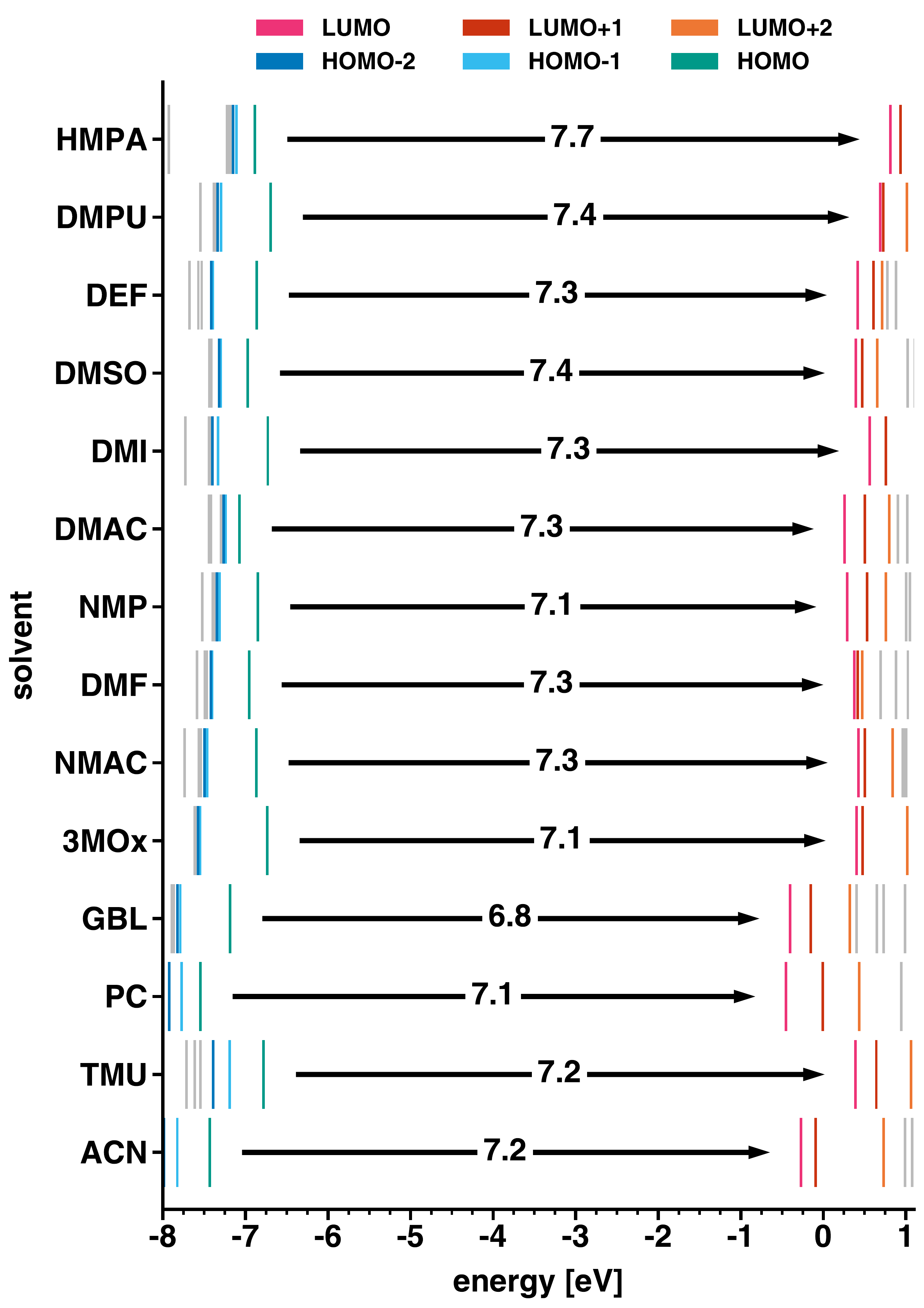}
    \caption{Energy distribution of the three highest-occupied and lowest-unoccupied molecular orbitals for the considered \ce{SnI2M4} complexes with $M$ being the solvent molecules listed on the $y$-axis. The HOMO-LUMO gaps are marked by the arrows and their values are reported in eV.}
    \label{fig:moenergies}
\end{figure}

It is worth deepening this analysis by inspecting the energy distribution of the MOs adjacent to the frontier (see Figure~\ref{fig:moenergies}).
In the majority of the considered complexes, the energetic separation between the HOMO and the HOMO-1 amounts to a few hundred meV.
In contrast, the energy difference between the HOMO-1 and the HOMO-2 is much smaller, never overcoming 50~meV.
A similar trend is found also for the \ce{PbI2M4} complexes~\cite{procida-etal2021firstprinciples}.
This situation is exacerbated in the tin-iodide complexes with DMPU, DMI, GBL, 3MOx, and NMAC, in which the separation between the HOMO and the HOMO-1 is close to 1~eV while the energies of HOMO-1 and HOMO-2 differ by less than 50~meV.
The spectrum of the unoccupied states is less regularly distributed in energy. 
The LUMO is generally closer to the LUMO+1 than the LUMO+1 to the LUMO+2, but no consistent trend can be identified.
Yet, there are some exceptions. 
In the complexes solvated with DMPU, DMSO, 3MOx, and NMAC, the energy difference between the LUMO and LUMO+1 is on the order of 50 meV while the higher unoccupied states lie at least 100~meV above the LUMO+1; in \ce{SnI2(NMP)4} and \ce{SnI2(PC)4}, the LUMO, LUMO+1, and LUMO+2 are almost equally separated at intervals of 230~meV and 440~meV, respectively; in the presence of HMPA and DMF, the eigenvalues of the first three unoccupied states are closely packed within a range of 100~meV.
It is important to underline that the orbital energies reported in Figure~\ref{fig:moenergies} are obtained from DFT+PCM and, as such, they cannot be interpreted as excitation energies in the framework of Koopman's theorem~\cite{krumland-etal2021exploring}. 
They only inform us about the distribution of the single-particle states in the electronic structure of the complexes, thus providing a useful starting point for the analysis of the optical properties below.

\begin{figure}[h]
    \centering
    \includegraphics[width=0.5\textwidth]{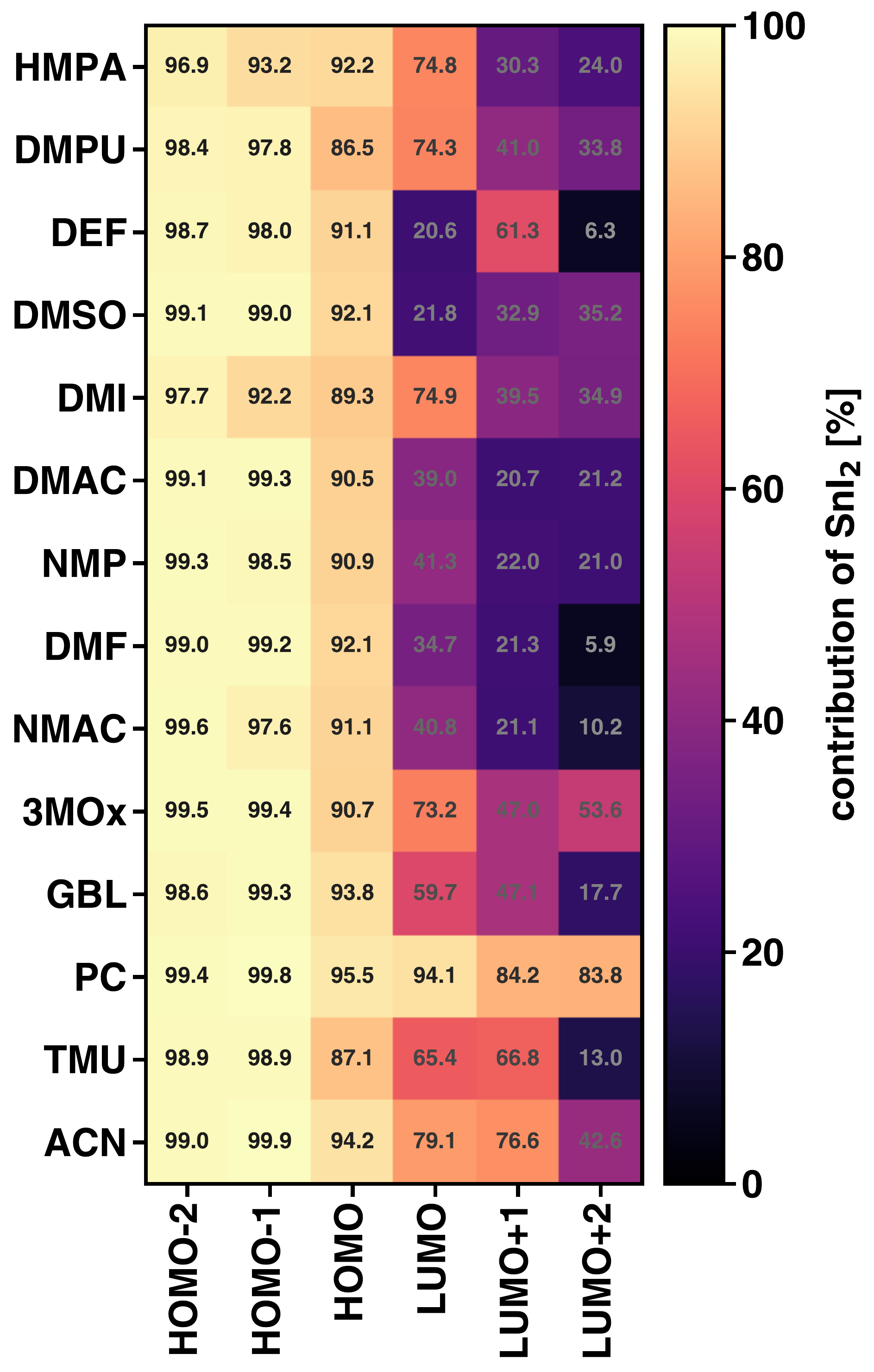}
    \caption{\ce{SnI2} contributions to the indicated molecular orbitals calculated with the NAO method. 
    }
    \label{fig:SnI2contribution}
\end{figure}

To complement the analysis of these energetic trends, we examine the spatial distribution of the MOs within the complexes.
For quantitative analysis, we make use of the NAO method and evaluate the \ce{SnI2} contribution to a certain orbital as the sum of the corresponding charge density on the Sn and I atoms (see Figure \ref{fig:SnI2contribution}).
The considered occupied states (HOMO-2, HOMO-1, and HOMO) are almost entirely localized on \ce{SnI2}.
In the HOMO-1 and HOMO-2, less than 10\% of the wave function is localized on the solvent, while in the HOMO it is below 15\%.
In the unoccupied region, the situation changes and substantial differences can be seen depending on the solvent.
However, in this case, no clear trends can be identified according to the donor number.
In the \ce{SnI2(PC)4} complex, the LUMO is localized by more than 94\% on the tin-iodide center. 
The presence of HMPA, DMPU, DMI, 3MOx, and ACN triggers a similar behavior with about 75\% of the orbital distributed on \ce{SnI2}.
In \ce{SnI2(GBL)4} and \ce{SnI2(TMU)4}, approximately 60\% and 65\% of the LUMO, respectively, sit on the metal-halide backbone.
In all other considered systems, the larger portion of the wave function of the lowest-unoccupied state is on the solvent molecules (see Figure~\ref{fig:SnI2contribution}).
At higher energies, the solvent contributions become even more prominent with the exception of the \ce{SnI2(PC)4} complex where all MOs analyzed in Figure~\ref{fig:SnI2contribution} are mainly focused on the \ce{SnI2} unit. 
Earlier results obtained for equivalent \ce{PbI2M4} systems indicate a similar trend, with the frontier states localized on the \ce{PbI2} kernel with negligible or even vanishing contributions from the surrounding solvent molecules~\cite{procida-etal2021firstprinciples,schier-etal2021formation}.
Further details on the orbital energies and spatial distributions of the inspected \ce{SnI2M4} complexes can be found in the Supporting Information, Figures~S1-S6. 


\subsection{Optical Properties}

\begin{figure}[h]
    \centering
    \includegraphics[width=\textwidth]{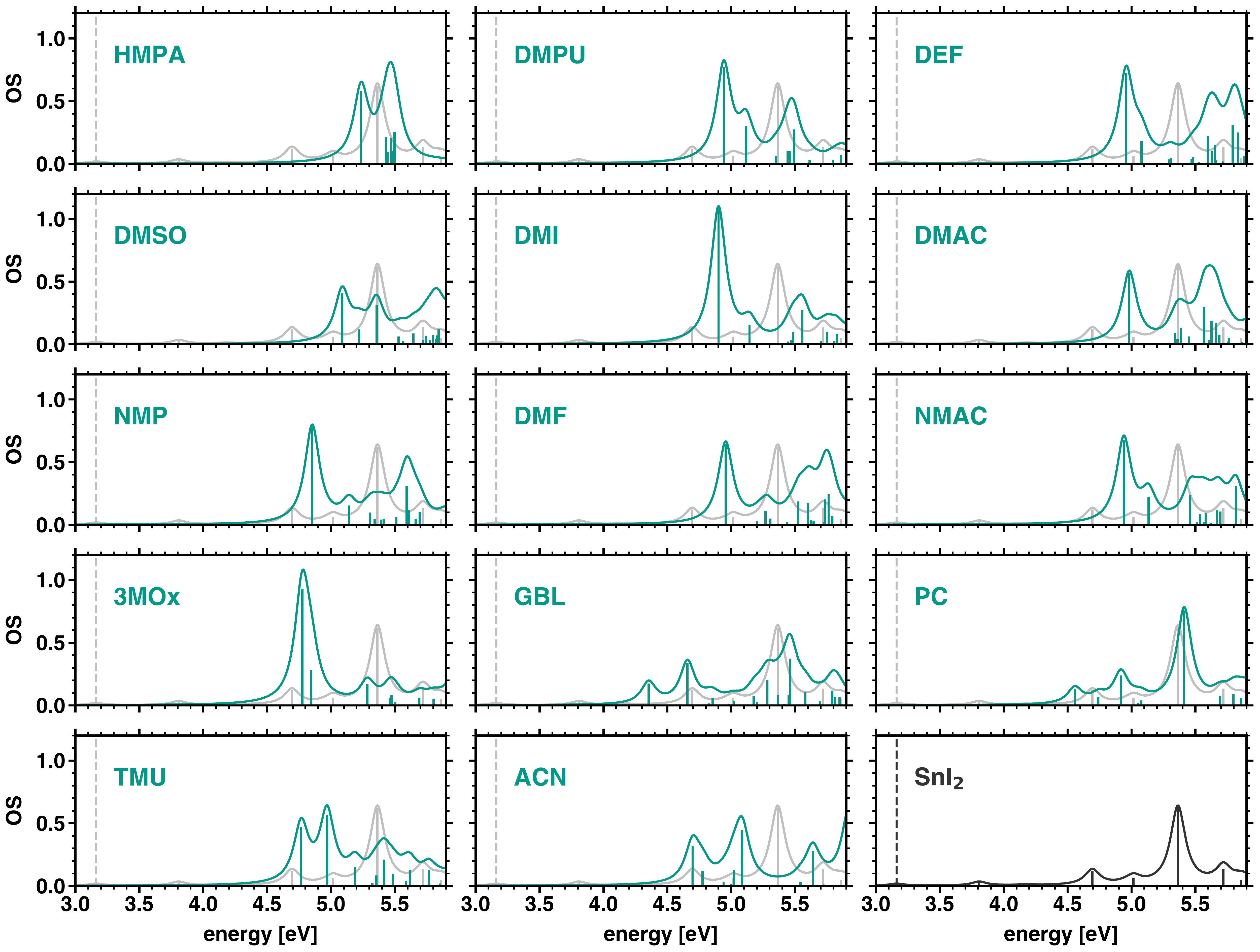}
    \caption{Optical absorption spectra of the 14 considered \ce{SnI2M4} complexes and of the \ce{SnI2} molecule in an implicit DMSO solution for reference. The absorption spectrum was calculated with a Lorentzian broadening of 70 meV.}
    \label{fig:absorptionspectra}
\end{figure}

The absorption spectra of the \ce{SnI2M4} complexes are reported in Figure~\ref{fig:absorptionspectra}, where the result obtained for the isolated \ce{SnI2} molecule in an implicit DMSO solution is displayed for comparison on the bottom-right panel.
We checked that the choice of the implicit solvent in this reference calculation does not impact the result obtained for the spectrum, as expected from previous analysis on other solvated compounds~\cite{krumland-etal2021exploring}.
The quantum-mechanical interactions with solvent molecules in the \ce{SnI2M4} compounds increase the energy of the first excited state from 3.2 eV (dashed bar in Figure~\ref{fig:absorptionspectra}) to 4~eV or higher (first peak in the green curves), overall redistributing the spectral weight toward lower energies. 
In fact, in the spectrum of \ce{SnI2}, the lowest-energy excitations have very weak intensity: the first peaks distinguishable in Figure~\ref{fig:absorptionspectra}, bottom-right panel, are above 4.5~eV and the most intense excitation in the displayed window is close to 5.4~eV.
Yet, weak but non-zero resonances are present at 3.1 and 3.8~eV.
In the majority of the spectra of the considered \ce{SnI2M4} complexes, on the other hand, the most intense resonance in the considered energy range is found below 5~eV.
Exceptions are the spectrum of \ce{SnI2(ACN)4}, which is dominated by two maxima of almost equal intensity at 4.7 and 5.1~eV, and by the results obtained for \ce{SnI2(DMSO)4} and \ce{SnI2(HMPA)4}, where the first bright transition is well above 5~eV. 
Notice that in the last two compounds, the HOMO-LUMO gap is the largest with values of 7.7~eV [\ce{SnI2(HMPA)4}] and 7.4~eV [\ce{SnI2(DMSO)4}], see Figure~\ref{fig:moenergies}.
The spectrum of \ce{SnI2(PC)4} is even more peculiar: it exhibits a remarkable similarity with the one of \ce{SnI2} alone.
This finding can be readily understood by recalling that both the HOMO and the LUMO in this solution complex are predominantly localized on the tin-iodide center with negligible wave-function delocalization on the solvent molecules (see Figure~\ref{fig:SnI2contribution}). 
A similar analysis performed on the subsequent four excitations (second to fifth excited states) in the spectrum of the complexes reveals a similar trend in terms of energy vs. donor number. 
However, it should be mentioned that these higher-lying excitations are dark in all systems with only a couple of exceptions (see Figure~S7).
This characteristic is due to the minimized wave-function overlap between the states involved in the corresponding transitions (see Figures~S8-S12), which occur between orbitals away from the frontier and, as such, localized on the different groups constituting the complexes (see Figure~\ref{fig:SnI2contribution}).

\begin{figure}[h]
    \centering
    \includegraphics[width=0.5\textwidth]{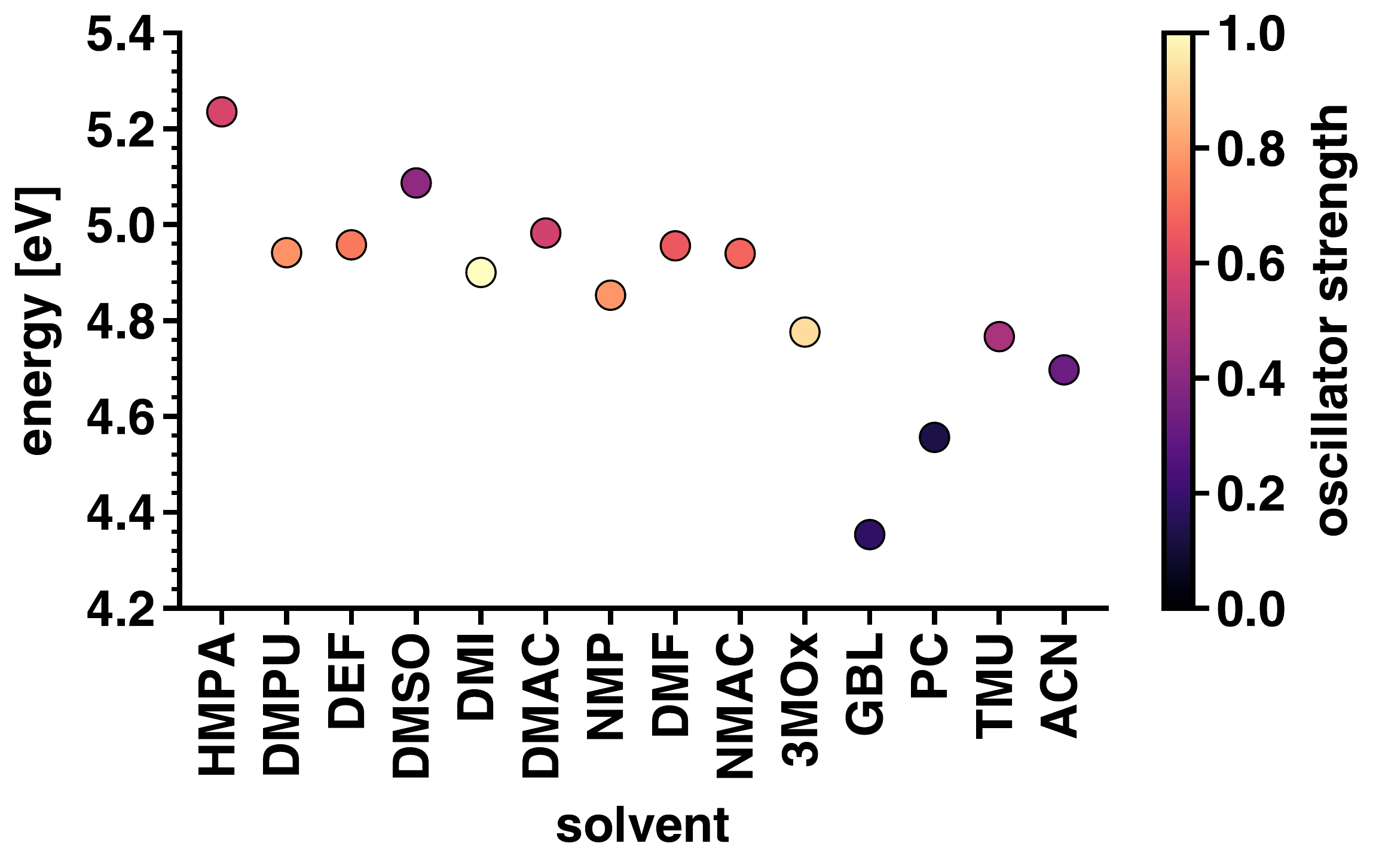}
    \caption{Energy and oscillator strength of the first excitation calculated for the 14 considered complexes. The solvent molecules on the $x$-axis are displayed with a decreasing donor number from left to right. 
    }
    \label{fig:firstexcitation}
\end{figure}

The characteristics of the first excitations in all considered complexes are summarized in Figure~\ref{fig:firstexcitation} (see also Table~S5).
In the graph correlating the energy and oscillator strength of the lowest-energy transition, we notice some trends.
First, there is a consistent (although not monotonic) red-shift of the first excitation with decreasing values of $D_N$, which is in general agreement with the measured UV-visible spectra of \ce{SnI2} solutions~\cite{cao-etal2021stability,heo-etal2021enhancing}.
Moreover, in accordance with physical intuition, the energy of the lowest-energy excitation correlates with the HOMO-LUMO gap.
Similar to the spectral weight, also this quantity decreases for decreasing values of $D_N$; moreover, the notable exception of \ce{SnI2(GBL)4} can be noticed as well in the trend of the HOMO-LUMO gaps in Figure~\ref{fig:h-l-gap}.

\begin{figure}[h]
    \centering
    \includegraphics[width=\textwidth]{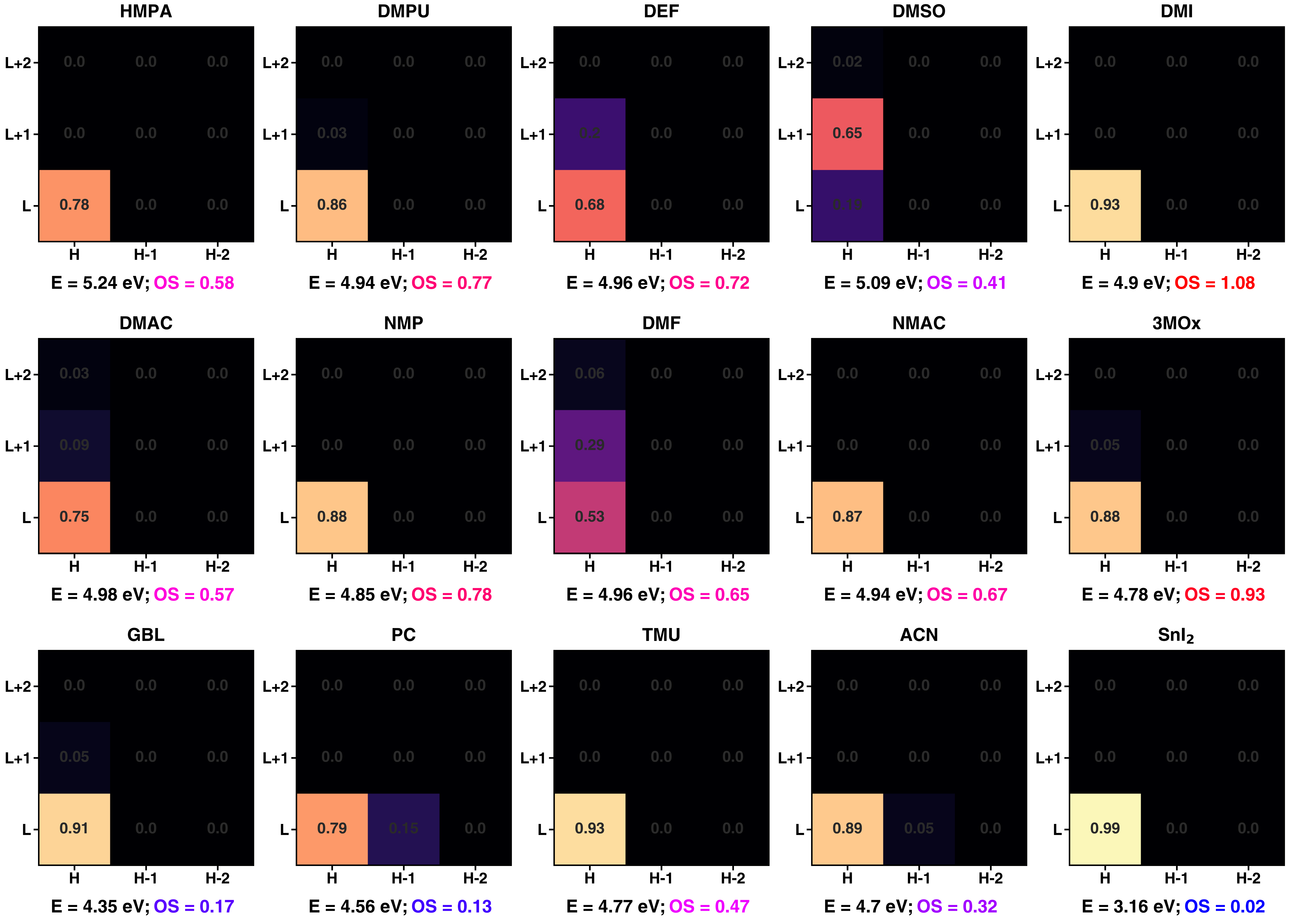}
    \caption{Composition of the first excited state of the 14 \ce{SnI2M4} complexes indicated by their solvent molecule $M$ and of the \ce{SnI2} molecule calculated in an implicit DMSO solution. The contribution $\in [0,1]$ of the transition from the occupied ($x$-axis) to the unoccupied state ($y$-axis) is displayed in the corresponding grid square. H stands for HOMO and L for LUMO. For each transition, the energy ($E$) and oscillator strength (OS) are reported. 
    }
    \label{fig:firstexccomposition}
\end{figure}

The low-energy optical transitions occurring in the complexes including solvents with high donor numbers are generally more intense than those in the spectra of low-$D_N$ solvents. 
This feature can be related to the orbital distribution reported in Figure~\ref{fig:SnI2contribution} through the analysis of the  excitations in terms of single-particle transitions, see Figure~\ref{fig:firstexccomposition}; in each box, the relative amount $\in [0,1]$ of each orbital transition is displayed together with the excitation energy and oscillator strength (OS). 
Not unexpectedly, the first excitation of almost all complexes stems primarily from the HOMO$\rightarrow$LUMO transition.
As such, the trend obtained for the HOMO-LUMO gaps (Figure~\ref{fig:h-l-gap}) is generally reflected in the first excitation energy (Figure~\ref{fig:firstexcitation}) as anticipated above.
An exception is given by \ce{SnI2(DMSO)4}, where the main contribution comes from a transition from the HOMO to the LUMO+1.
A likely reason for the atypical result, which corresponds to comparably lower measured absorption of Sn-perovskite films with DMSO compared to other solvents~\cite{heo-etal2021enhancing}, can be related to the larger difference between the spatial distribution of HOMO and LUMO in \ce{SnI2(DMSO)4}.
While, like in all complexes, the HOMO is almost entirely formed by $s$- and $p$-states of Sn and I atoms (see Figures~S3 and S6), the wave function of the LUMO is mainly distributed on the solvent atoms.
This hypothesis is supported by the fact the two other complexes with lower HOMO$\rightarrow$LUMO contribution to the first excitation, namely \ce{SnI2(DEF)4} and \ce{SnI2(DMF)4}, have also very low \ce{SnI2} contributions to the LUMO with 20.6\% and 34.7\%, respectively (see Figure~\ref{fig:SnI2contribution}).
However, contrary to the results obtained for the \ce{PbI2M4} complexes~\cite{procida-etal2021firstprinciples}, a low HOMO$\rightarrow$LUMO contribution does not necessarily correspond to a low OS of the first excitation in the \ce{SnI2M4} systems.
In fact, while the interaction with solvent molecules causes a significant reduction of the OS due to charge delocalization away from the \ce{PbI2} unit in the Pb-based complexes, this is not the case in the Sn-based systems.

\section{Summary and Conclusions}
In summary, we have investigated the structural, energetic, electronic, and optical properties of 14 tin-iodide solution complexes. 
As model systems, we have considered complexes with formula \ce{SnI2M4}, where $M$ are common solvents with complementary chemical characteristics and varying donor numbers.
With this approach, we were able to model the short-range quantum-mechanical interactions between solute and solvents that are expected to dominate the electronic and optical response of the systems.
All the investigated structures are stable although, in two compounds, not all solvent molecules form chemical bonds with \ce{SnI2}. 
The formation energy, introduced as a metric for stability, decreases in magnitude with decreasing values of $D_N$, suggesting that the ability of the solvents to bind to tin iodide depends directly on their effectiveness in donating electrons. 
The orbital energies do not follow an equally clear trend, although solvents with higher donor numbers tend to have higher energies for the frontier orbitals and also for the gap between them. 
The energetic distribution of the highest occupied and lowest unoccupied states could not be straightforwardly related to $D_N$ either, although their spatial distribution follows a clear trend. 
The occupied orbitals are predominantly localized on the \ce{SnI2} unit while the unoccupied ones are, in most cases, largely distributed also on the solvent molecules. 
The limited wave-function overlap between the frontier states is responsible for the low oscillator strength of the first excitation in most complexes. 
This being said, the presence of the coordinated solvent molecules red-shifts the spectral weight in comparison to the result obtained for \ce{SnI2} alone.

To conclude, our results provide a comprehensive characterization of the fundamental properties of tin-iodide solution complexes. 
Thanks to the fully quantum-mechanical approach adopted in this study and the deep level of our analysis, our findings complement the existing knowledge available for these systems offering a robust reference for data analysis, interpretation, and understanding. 
Moreover, they can be the basis for future investigations on halide perovskites solution precursors and/or compounds produced in subsequent synthesis steps of tin-halide perovskite thin films.  

\begin{acknowledgement}
We are grateful to Mahmoud H. Aldamasy for sharing valuable information on the solvent molecules investigated in this study.
We thank Antonio Abate for inspiring discussions and for critical reading of the unpublished version of this manuscript.
This work was supported by the German Research Foundation through the Priority Program SPP 2196 (Project number 424394788), by the German Federal Ministry of Education and Research (Professorinnenprogramm III), and from the State of Lower Saxony (Professorinnen f\"ur Niedersachsen and Nieders\"achsisches Vorab -- project SMART).

\end{acknowledgement}

\begin{suppinfo}
Additional details on the \textit{ab initio} calculation of Gutmann's donor number are provided~\cite{miranda-quintana-smiatek2021calculation,laurence-etal2011overview,smiatek2019enthalpic} together with the data related to structural properties, orbital energies, and spatial distribution, and to the analysis of the optical excitations.

\end{suppinfo}

\section*{Data Availability Statement}
The data that support the findings of this study are openly available in
Zenodo at DOI 10.5281/zenodo.7729359.


\providecommand{\latin}[1]{#1}
\makeatletter
\providecommand{\doi}
  {\begingroup\let\do\@makeother\dospecials
  \catcode`\{=1 \catcode`\}=2 \doi@aux}
\providecommand{\doi@aux}[1]{\endgroup\texttt{#1}}
\makeatother
\providecommand*\mcitethebibliography{\thebibliography}
\csname @ifundefined\endcsname{endmcitethebibliography}
  {\let\endmcitethebibliography\endthebibliography}{}

\end{document}